\documentclass[aps,pra,twocolumn,superscriptaddress,10pt,a4paper,reprint,nofootinbib]{revtex4-1}
\usepackage{amsmath,amssymb,amsfonts}
\usepackage{mathtools}
\usepackage{graphicx}
\usepackage{mwe}
\usepackage{txfonts}
\usepackage{color}
\usepackage{physics}
\usepackage{mhchem}

\newcommand{\documenttitle}{Out of Equilibrium Behaviour of Quantum Vortices: A Comparison of Point Vortex Dynamics and Fokker-Planck Evolution}
\newcommand{\documentauthor}{R. J. Tattersall}

\usepackage[unicode,breaklinks]{hyperref}
\hypersetup{
	unicode=true,
	a4paper=true,
	plainpages=false,
	pdftitle={\documenttitle},
	pdfauthor={\documentauthor},
	pdfsubject={\documenttitle},
	colorlinks=true,
	linkcolor=blue,
	citecolor=blue,
	filecolor=black,
	urlcolor=blue
}
\begin{document}
	\title{\documenttitle}
	\author{\documentauthor}
	\affiliation{Joint Quantum Centre (JQC) Durham-Newcastle, School of Mathematics Statistics and Physics, Newcastle University, Newcastle upon Tyne, NE1 7RU, UK}
	\author{A. W. Baggaley}
	\affiliation{Joint Quantum Centre (JQC) Durham-Newcastle, School of Mathematics Statistics and Physics, Newcastle University, Newcastle upon Tyne, NE1 7RU, UK}
	\author{T. P. Billam}
	\affiliation{Joint Quantum Centre (JQC) Durham-Newcastle, School of Mathematics Statistics and Physics, Newcastle University, Newcastle upon Tyne, NE1 7RU, UK}
	
	\date{\today}

\begin{abstract}
An out of equilibrium two-dimensional superfluid relaxes towards equilibrium via a process of coarsening, driven by the annihilation of vortices with antivortices.  Here we present a comparison of two different numerical models of this process, a dissipative point vortex model and Fokker-Planck evolution, across a wide range of initial configurations and levels of dissipation.  We find that our dissipative point vortex model is well-approximated by Fokker-Planck evolution only for very low initial energies per vortex, $E^{0}/N_{\varv}\lesssim -4$, when almost all vortices and antivortices are closely bound into dipoles.  We observe that the dynamical critical exponent, $z$, in the dissipative point vortex model, undergoes a crossover, from a roughly constant value close to two for $E^{0}/N_{\varv}\gtrsim -4$ to one which depends explicitly on the initial conditions for $E^{0}/N_{\varv}\lesssim -5$.
\end{abstract}

\maketitle

\section{\label{sec:intro}Introduction} 
When a system undergoes a rapid quench from a disordered to an ordered phase, it does not order instantly but instead relaxes towards equilibrium over time.  During this relaxation, the dynamical scaling hypothesis predicts that the length scale of ordered regions increases, with later configurations statistically similar to earlier ones except for a change in global scale, characterised by a correlation length $L_{c}(t)$ \cite{Bray1994}.  Such phase-ordering kinetics is observed in a wide range of physical systems, including the Ising and XY magnetic models \cite{Humayun1991,Rojas1999,Jelic2011,Yurke1993}, decomposition of binary alloys \cite{Gaulin1987} and the non-equilibrium behaviour of quantum fluids \cite{Groszek2021}.  Furthermore, the concept of universality suggests that this scaling behaviour may be characterized by a limited number of exponents that reflect the dimensionality of the system and relevant conservation laws but are independent of the microscopic Hamiltonian \cite{Hohenberg1977}.  In this paper we focus on the coarsening behaviour of a two-dimensional superfluid following a quench and compare the results of two reduced models of the system across a range of initial conditions. 

Following the creation of Bose-Einstein condensates (BECs) in ultracold atomic gases in the mid-1990s \cite{Anderson1995,Davis1995} and accelerated by more recent developments in imaging and manipulation of the atomic condensates \cite{Abo-Shaeer2002,Lin2009,Wilson2015,Gauthier2016}, the non-equilibrium dynamics of two-dimensional superfluids has become an active area of experimental research.  Studies include observations of scaling in the momentum distribution of a one-dimensional BEC \cite{Erne2018} and in the spatial correlations of spin excitations in a quasi-one-dimensional spinor BEC close to non-thermal fixed points \cite{Prufer2018}.  There have also been signs of scaling behaviour observed in vortex number decay for large scale clusters formed in two-dimensional BECs trapped in highly oblate geometries \cite{Johnstone2019}.  Experiments using three-dimensional BECs have also demonstrated universal scaling in the momentum distribution far from equilibrium \cite{Glidden2021,Garcia_Orozco2022}.  

There is also a rich history of numerical investigations into coarsening behaviour of a two-dimensional superfluid.  Several recent studies have applied techniques based on the Gross-Pitaevskii equation (GPE) \cite{Pethick2008,Pitaevskii2016}, in which the fluid is described by a single wavefunction, $\psi$, and scaling behaviour is observed in the growth of spatial correlations in the phase following a quench \cite{Karl2017,Groszek2020,Groszek2021}.  Many studies have focussed instead on the closely related planar XY model.  In these, it is correlations between the angle of magnetic spins which demonstrate scaling over time \cite{Rojas1999,Jelic2011,Yurke1993}.  In both GPE and XY models, vortices play a key role in the coarsening dynamics.  In the GPE these are regions of zero density around which the phase has a wrapping of $+2\pi$ for a vortex or $-2\pi$ for an antivortex, whilst in the XY model these are points around which the surrounding magnetic spins undergo a change in direction of $\pm 2\pi$ for one circuit.  Dissipation tends to drive vortices and antivortices together, whereupon they annihilate.  These annihilations drive the growth in regions of constant phase/angle, increasing the correlation length of the order parameter.  

Numerical modelling of two-dimensional Bose gases has generally concluded that the dynamical scaling constant $z$, which characterises the growth of the correlation length according to $L_{c}(t)\sim t^{1/z}$, is close to two when the microscopic dynamics includes some dissipation \cite{Groszek2021}.  We note that it has been suggested that in some cases a logarithmic correction to the scaling equation of the form $L_{c}(t)\sim (t/\ln{t})^{1/z}$ is required \cite{Bray2000}.  This is in accordance with the Model A dynamical universality class \cite{Hohenberg1977} for a two-dimensional system with non-conserved order parameter. This type of universal scaling would be expected to hold regardless of the choice of, sufficiently non-equilibrium, initial conditions.

However, an alternative prediction has been made for quenches from below the Kosterlitz-Thouless (KT) transition.  The KT transition in two-dimensional superfluids is a topological phase transition driven by vortex-antivortex unbinding \cite{Kosterlitz1973,Kosterlitz1974} and has been observed, for example, in superfluid helium films \cite{Bishop1978,Bishop1980}.  Below the critical temperature, vortices and antivortices are bound in dipole pairs and the time-dependence of the distribution of dipole lengths has been modelled using a Fokker-Planck equation \cite{Chu2000,Chu2001,Forrester2013,Williams2022}.  This equation simulates the drift of dipoles towards shorter lengths, caused by dissipation, as well as thermally driven diffusion.  A cut-off at one healing length takes the place of vortex-antivortex annihilation.  Numerical and analytical studies using this method have found scaling behaviour with a value for $z$ that depends explicitly on the temperature prior to the quench and can differ markedly from two \cite{Chu2001,Forrester2013,Williams2022}.  This change to the scaling laws may be a consequence of the long-ranged (power-law) spatial correlations below the KT transition \cite{Bray1991,Bray1994}.   

\begin{figure}
	\centering
	\includegraphics[width=8.7cm]{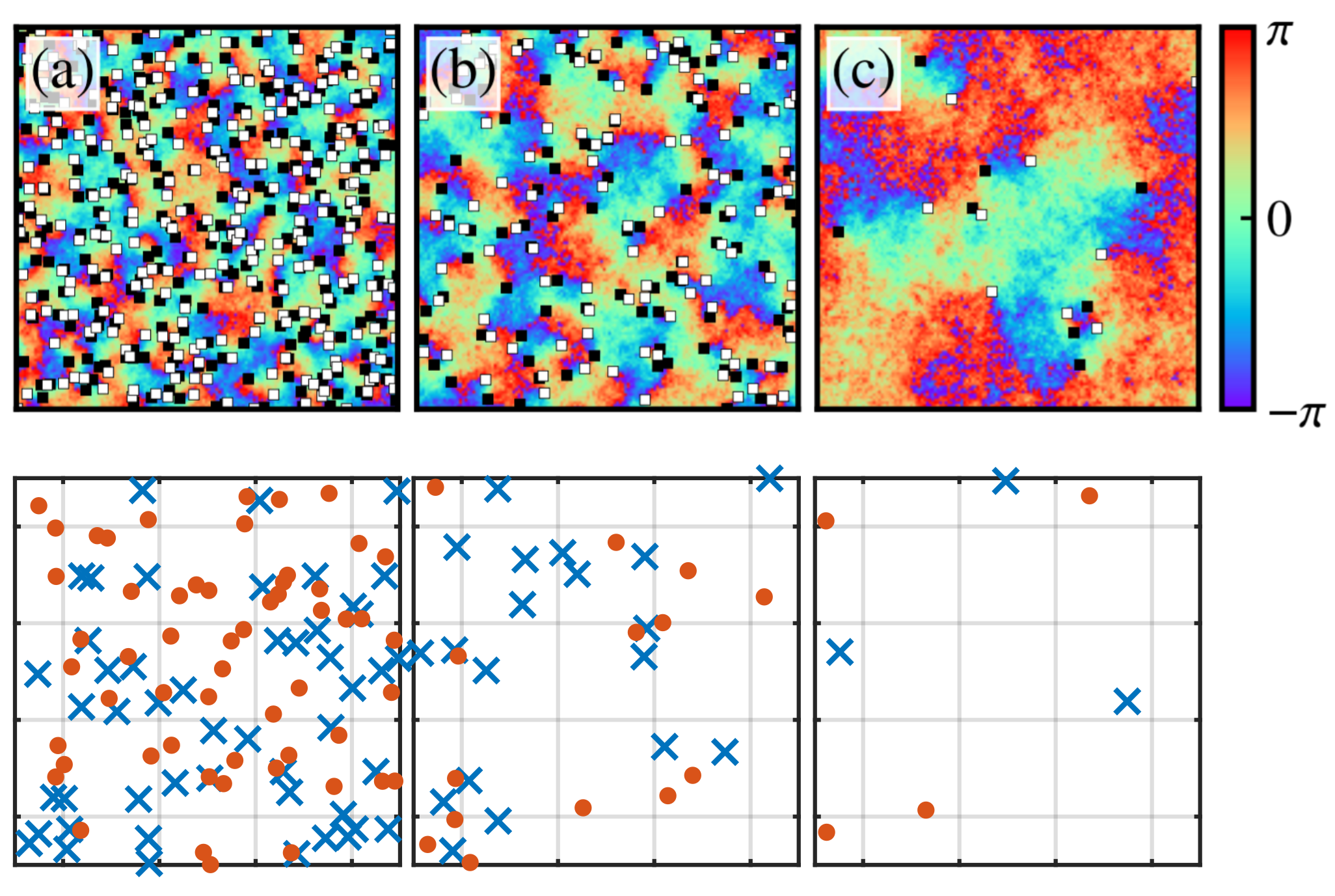}
	\caption{Coarsening of a two-dimensional superfluid following a quench.  The top row shows the results of one Gross-Pitaevskii simulation of the field $\psi$ of a BEC in the case of zero dissipation, reproduced from \cite{Groszek2021} under CC-BY-4.0 international license (https://creativecommons.org/licenses/by/4.0/).  The open white squares show the positions of vortices (positive circulation), the filled black squares show the positions of antivortices (negative circulation) and the background colour indicates the quantum mechanical phase.  For a qualitative comparison, the bottom row shows snapshots from one evolution of the dissipative point vortex model used in this paper.  Blue crosses mark the positions of vortices and filled red circles mark antivortices.  Both rows show time progressing from left to right and in both models the annihilation of vortices and antivortices is the key mechanism driving the coarsening process.}\label{fig:coarsening}
\end{figure}

The basic objects of the Fokker-Planck approach are vortex dipoles and it provides an effective model of two-dimensional superfluids at large scales, whilst not including compressible excitations such as phonons.  An analogous, phonon-free, microscopic model is the dissipative point vortex model, in which simulating the fluid is reduced to calculating the motions and interactions of $\delta$-function like vortices and antivortices  \cite{Helmholtz1858,Kirchoff1876}.  The dynamics of vortices in the GPE can be mapped onto the point vortex model \cite{Fetter1966}, so long as the separation of adjacent vortices is much larger than the vortex core size, and a few studies have also observed evidence of scaling behaviour using this simple model \cite{Huber1993,Salman2016,Tattersall2024}.  See Fig.~\ref{fig:coarsening} for an illustration of phase ordering as seen in both GPE simulations (top row) \cite{Groszek2021} and in simulation data found using the dissipative point vortex model employed in this paper.

In this paper we compare the results of simulating an out of equilibrium two-dimensional superfluid using a dissipative point vortex model with the results of a Fokker-Planck equation for dipole lengths.  We do this across a range of initial conditions and for different levels of dissipation.  The  point vortex model used in this paper does not account explicitly for thermal effects; nevertheless, the inclusion of dissipation ensures that vortices and antivortices are drawn together and, by removing any that get closer than one healing length, mimics the annihilations which drive coarsening.  To test the effect of changing initial conditions we make use of the energy per vortex.  This depends on the spatial configuration of the vortices and antivortices and provides a useful characterisation of the initial conditions \cite{Weiss1991}.  For low values, the vortices and antivortices become closely bound into dipoles, whereas at higher values they are arranged more randomly.  At higher energies still they would form clusters of either vortices or antivortices. We find that for the very lowest initial energies per vortex, the dipole length distributions extracted from point vortex simulations show good agreement with those found using the Fokker-Planck equation.  For the same low energy initial conditions, we also find evidence of a change in scaling behaviour in the point vortex results, from an apparently universal value of $z\approx 2$ at higher energies to one which depends explicitly on the initial configuration as predicted in \cite{Forrester2013}.

The structure of this paper is as follows.  In section~\ref{sec:background} we outline the theoretical backgound to the dynamical scaling hypothesis, the point vortex model and the Fokker-Planck equation as applied to two-dimensional superfluids.  In section~\ref{sec:methodology} we decribe the numerical methodology for creating initial configurations of vortices and antivortices and compare them with expectations for a system of widely-spaced, independent dipoles.  We also simulate their dynamics using a dissipative point vortex model and calculate the evolution of the dipole length distribution using a Fokker-Planck equation.  In section~\ref{sec:results} we present results comparing the point vortex and Fokker-Planck methods.  We also describe an analysis of the dynamical scaling constant, $z$, which shows evidence of a change in behaviour at low initial energies per vortex.  In section~\ref{sec:conclusions} we summarise our conclusions.

\section{\label{sec:background}Background}
\subsection{\label{subsec:scaling_hypothesis}Scaling Hypothesis} 
For a two-dimensional system undergoing a phase-ordering process, the scaling hypothesis \cite{Bray1994} predicts that the correlation function, $G(r,t)$, between two points separated by a distance $r$, is a scaling function.  This means that changes in time, $t$, can be replaced with a rescaling in length by $L_{c}(t)$, where $L_{c}(t)$ is a correlation length that depends on time only:
\begin{equation}
G(r,t)=G(r/L_{c}(t))\,.
\label{eqn:scaling}
\end{equation}
Furthermore, when dynamical scaling holds, the dependence of correlation length on time is predicted to be a power law of the form:
\begin{equation}
L_{c}\sim t^{1/z}\,,
\label{eqn:corr_length_power_law}
\end{equation}
where $z$ is the dynamical critical exponent.  The relevant dynamical universality class for a two-dimensional quantum fluid, with dissipation, is argued to be Model A \cite{Hohenberg1977,Halperin1972}, with $z=2$, resulting in $L_{c}\sim t^{1/2}$.  If we make the assumption that the typical area per vortex $\sim L_{c}^{2}$, it follows that the number of vortices, $N_{\varv}$, should follow:
\begin{align}
N_{\varv}&\sim t^{-2/z}\,,\label{eqn:Nv_power_law}\\
\rightarrow N_{\varv}&\sim t^{-1}\,\text{(for $z=2$)}. \nonumber
\end{align}

Logarithmic corrections have been proposed to the form of these scaling laws within the context of the planar XY model.  One suggestion is that in the presence of free vortices the relation $L_{c}\sim t^{1/2}$ should be replaced with $L_{c}\sim (t/\ln{t})^{1/2}$ \cite{Rutenberg1995,Bray2000}, whilst another argues that the density of vortices, $\rho$, should follow $\rho\ln{\rho}\sim t^{-1}$ instead of $\rho\sim t^{-1}$ \cite{Yurke1993}.  However, comparison of fitting both a corrected and an uncorrected power law to data from GPE modelling \cite{Groszek2021} yielded equally close fits in both cases, but with slightly shifted values for $z$.  In this paper we apply the uncorrected Eq.~\ref{eqn:corr_length_power_law} when analysing scaling behaviour in our point vortex results.

\subsection{\label{subsec:pvm}Point Vortex Model Description}
The point vortex model is a classical model of fluid dynamics in two dimensions in which the vorticity is assumed to be concentrated in $\delta$-function like points and the fluid circulates around these points.  Points with positive circulation are vortices and points with negative circulation are antivortices \cite{Helmholtz1858,Kirchoff1876}.  The dynamics of the fluid is then reduced to finding the motions of these vortices \cite{Newton2001}, each of which is advected by the local fluid velocity.  The equations of motion for vortex $j$ in an infinite plane are therefore:
\begin{align}
\dv{x_j}{t} &=-\frac{1}{2\pi}\sum_{i=1,i\neq j}^{N}\frac{\kappa_i(y_{j}-y_{i})}{|\mathbf{r}_{j}-\mathbf{r}_{i}|^2}\, , \nonumber\\
\dv{y_j}{t} &=\frac{1}{2\pi}\sum_{i=1,i\neq j}^{N}\frac{\kappa_i(x_{j}-x_{i})}{|\mathbf{r}_{j}-\mathbf{r}_{i}|^2}\, ,
\label{eqn:equationsofmotion}
\end{align} 
in which $\mathbf{r}_{j}=(x_{j},y_{j})$ are the coordinates of vortex $j$, $\mathbf{r}_{i}=(x_{i},y_{i})$ are the coordinates of all of the other vortices, and $\kappa_{i}$ are the circulations of these other vortices.  Note that for quantum fluids all circulations are either $+h/m$ or $-h/m$, where $h$ is the Planck constant and $m$ is the mass of an atom.  The Hamiltonian for this system can be written as \cite{Novikov1975}:
\begin{equation}
H=-\frac{1}{4\pi} \sum_{i=1,i\neq j}\kappa_i\kappa_j \ln{|\mathbf{r}_{i}-\mathbf{r}_{j}|}\, . \label{eqn:Hamiltonian}
\end{equation}  
The point vortex model does not include compressible effects, such as vortex-sound interactions, it also ignores the structure of the vortex cores; however, if the separation of the vortices is much greater than the core size, it provides a good model of vortex dynamics \cite{Fetter1966}.  Futhermore, by making calculations based on just vortex coordinates and signs and not on the wavefunction for the whole fluid, its relative computational simplicity compared to modelling with the Gross-Pitaevskii equation makes it attractive for simulating large systems.  In order to mimic annihilation of vortices with antivortices, we simply remove any vortex-antivortex pair that come closer than one healing length, $\xi$.  This is the characteristic length for a superfluid and is the distance over which the density recovers from zero to the bulk value; hence it is a good estimate of the radius of a vortex core.  By adding annihilations in this manner, we introduce into our model the most important, qualitative feature of vortex-sound interactions.

For our simulations, we consider vortices in a doubly-periodic square box of length $L$.  The corresponding equations of motion and Hamiltonian then become \cite{Weiss1991}:
\begin{align}
\frac{d{\tilde{x}_{i}}}{dt}&=\sum_{j=1,j\neq i}^{N_{v}} \kappa_j\sum_{m=-\infty}^{\infty}  \frac{-\sin(\tilde{y}_{ij})}{\cosh(\tilde{x}_{ij}-2\pi m)-\cos(\tilde{y}_{ij})}\, , \nonumber\\
\frac{d{\tilde{y}_{i}}}{dt}&=\sum_{j=1,j\neq i}^{N_{v}} \kappa_j\sum_{m=-\infty}^{\infty}  \frac{\sin(\tilde{x}_{ij})}{\cosh(\tilde{y}_{ij}-2\pi m)-\cos(\tilde{x}_{ij})}\, .
\label{eqn:pv_eqm_periodic}
\end{align}
\begin{equation}
H_{\square}=-\sum_{i=1}^{N_{\varv}}\sum_{j=1,j \neq i}^{N_{\varv}} \frac{\kappa_i \kappa_j}{2}\sum_{m=-\infty}^{\infty} \ln{\left( \frac{\cosh{\tilde{x}_{ij}-2\pi m}-\cos{\tilde{y}_{ij}}}{\cosh{2\pi m}}\right)}-\frac{\tilde{x}_{ij}^2}{2\pi}\, ,
\label{eqn:pv_H_periodic}
\end{equation}
where $\kappa_{i}$ is the circulation of vortex $i$ and its coordinates, $(x_{i},y_{i})$, are rescaled by $2\pi/L$, to give $(\tilde{x}_{i},\tilde{y}_{i})$.  Similarly, vortices $j$ have circulations $\kappa_{j}$ and their coordinates $(x_{j},y_{j})$ are rescaled to $(\tilde{x}_{j},\tilde{y}_{j})$, so that in these units the box has a width of $2\pi$.  $\tilde{x}_{i}-\tilde{x}_{j}$ and $\tilde{y}_{i}-\tilde{y}_{j}$ are written as $\tilde{x}_{ij}$ and $\tilde{y}_{ij}$.  The time units are also scaled to give $\kappa_{i}$ and $\kappa_{j}=\pm 1$.  In all of these equations the term in the second sum falls rapidly with increasing $m$, therefore an excellent level of precision is achieved by truncating the limits of this sum from $m=\pm\infty$ to $m=\pm 5$.

\subsection{\label{subsec:FPT}Fokker-Planck Description}
Below the Kosterlitz-Thouless transition \cite{Kosterlitz1973,Kosterlitz1974}, vortices are bound in vortex-antivortex pairs known as dipoles.  If $\Gamma$ is the probability distribution of the dipole lengths, then its evolution over time may be subject to modelling using a Fokker-Planck equation \cite{Gardiner2009}, as explained in detail in Ref.~\cite{Forrester_Dissertation}.  The general form of which, for a probability distribution $\Gamma(\mathbf{R},t)$, where $\mathbf{R}$ is a vector containing the state space variables, is:
\begin{equation}
\pdv{\Gamma}{t}=-\pdv{}{R_{i}}\left(M_{1}^{i}\Gamma\right)+\pdv[2]{}{R_{j}}{R_{k}}\left(M_{2}^{jk}\Gamma\right)\,.
\label{eqn:general_FP}
\end{equation}
$M_{1}^{i}(\mathbf{R},t)$ is called the \textit{drift vector} and is given by:
\begin{equation*}
M_{1}^{i}(\mathbf{R},t)=\lim_{\Delta t\to 0}\frac{\langle \Delta R_{i}\rangle(\mathbf{R},t)}{\Delta t}\,,
\end{equation*}
in which the $\langle \Delta R_{i}\rangle$ is the change in the $i^{\text{th}}$ component of $\mathbf{R}$ in a time $\Delta t$, averaged over all of the particles being modelled; in our case these are the dipoles.

The matrix $M_{2}^{jk}(\mathbf{R},t)$ is the  called the \textit{diffusion matrix} and is given by:
\begin{equation*}
M_{2}^{jk}(\mathbf{R},t)=\lim_{\Delta t\to 0}\frac{\langle \Delta R_{i} \Delta R_{k}\rangle(\mathbf{R},t)}{2\Delta t}\,.
\end{equation*}  

For a two-dimensional superfluid below the KT transition, the only relevant variable is the dipole length, $l$.  Therefore, by making the substitution $\mathbf{R}\rightarrow l$ it is possible to derive a one-dimensional form of the Fokker-Planck equation \cite{Ambegaokar1980,Chu2001,Forrester2013,Forrester_Dissertation,Williams2022}:
\begin{equation}
\pdv{\Gamma}{t}= \left(\frac{1+2\pi K}{l}\right)\pdv{\Gamma}{l}+\pdv[2]{\Gamma}{l}\,.
\label{eqn:FP_dipoles_Forrester}
\end{equation}
In this form, $t$ is the time measured in units of the diffusion time and $l$ is the dipole length measured in units of the healing length, $\xi$.  $K$ is the dimensionless superfluid density given by:
\begin{equation*}
K=\frac{\hbar^{2}\sigma_{s}}{m^{2}k_{B}T}\,,
\end{equation*}
where $\hbar$ is the reduced Planck constant, $\sigma_{s}$ is the dimensionful superfluid density, $m$ is the atomic mass, $k_{B}$ is the Boltzmann constant and $T$ is the temperature \cite{Nelson1977}. 

The first term in Eq.~\ref{eqn:FP_dipoles_Forrester} corresponds to a \textit{drift} towards smaller dipole lengths and is related to dissipation, whilst the second represents a random \textit{diffusion} in dipole lengths.  Whilst diffusion is thermally driven in systems such as the XY model, our point vortex simulations do not explicitly include thermal effects.  Nevertheless, we anticipate that the effect of many small interactions between different dipoles, as they move relative to each other, could manifest in similar stochastic changes in the dipole lengths.  

Since one healing length is the distance at which a vortex and antivortex annihilate, we can write the normalization condition:
\begin{equation}
\int_{1}^{\infty}\Gamma\, 2\pi ldl = \rho = \frac{N_{d}}{L^{2}}\,,
\label{eqn:Forrester_normalization}
\end{equation}
where $N_{d}$ is the number of dipoles, $L$ is the width of the box and $\rho$ is the dipole density (number of dipoles per unit area) in the system.  The density, $\rho$, reduces in time as dipoles shrink to one healing length and their constituent vortices and antivortices annihilate.  

In equilibrium we require $\partial\Gamma/\partial t =0$ and so, if we have a collection of dipoles at equilibrium for some initial superfluid density $K_{i}$, it follows from Eq.~\ref{eqn:FP_dipoles_Forrester} that the distribution of lengths must be of the form:
\begin{equation}
\Gamma\sim l^{-2\pi K_{i}}\,.
\label{eqn:initial_gamma}
\end{equation}
In Ref.~\cite{Forrester2013} an exact analytic solution to Eq.~\ref{eqn:FP_dipoles_Forrester} is derived for a quench from a constant initial superfluid density, $K_{i}$, to a constant final density, $K_{f}$.  After a short time, this solution gives the time dependence of the dipole density as:
\begin{equation*}
\rho(t)\sim t^{(1-\pi K_{i})}\,.
\label{eqn:Forrester_rho_t}
\end{equation*}
Since the number of dipoles is simply half the number of vortices, it follows from Eq.~\ref{eqn:Nv_power_law} that:
\begin{equation*}
\rho(t)\sim t^{-2/z}\,.
\end{equation*}
Comparison of these two expressions for $\rho (t)$ yields the following prediction for the dynamical scaling constant for quenches below the KT transition:
\begin{equation}
z=\frac{2}{(\pi K_{i}-1)}\,.
\label{eqn:FP_z}
\end{equation}
This result depends explicitly on the initial conditions of the quench and differs from the Model A prediction, $z= 2$.

\section{\label{sec:methodology}Methodology}
\subsection{\label{subsec:initial_conditions}Creating Thermalized Initial Conditions at Different Vortex Energies}
Within the point vortex model the only energy is the incompressible kinetic energy of the fluid.  This is entirely determined by the positions and circulations of the vortices through Eq.~\ref{eqn:pv_H_periodic}.  Scaling by the number of vortices, $N_{\varv}$, gives the energy per vortex, $E/N_{\varv}$;  when generating initial conditions, this is a useful parameter.  Lowering $E/N_{\varv}$ corresponds to moving vortices and antivortices closer together to form smaller dipoles and increasing it tends to form clusters of either only vortices or only antivortices.  Consequently, two configurations with the same energy per vortex share similar levels of clustering and/or similar distributions of dipole lengths.  Our aim is to create an ensemble of such configurations that is in microcanonical thermal equilibrium at the desired $E/N_{\varv}$, in the context of the point vortex model, subject to the constraint that the minimum separation of the vortices is $\xi$ and so there are no annihilations.

To create these initial conditions, with specified values of $E/N_{\varv}$, we follow an approach similar to \cite{Billam2014,Groszek2018a} by first scattering $2069$ vortices and $2069$ antivortices in a doubly-periodic square box of length $L=2048$ (measured in units of $\xi$).  These numbers were chosen to give a similar initial vortex density to those found in the GPE simulations shown in the top row of Fig.~\ref{fig:coarsening}, although our much larger domain size ($2048\xi$ compared to $363\xi$) increases the robustness of our later statistical analysis.  We then employ a biased random walk algorithm, in which the positions of vortices are updated at random and the updated positions accepted only if they bring $E/N_{\varv}$ closer to the desired value.  Once the desired value is achieved to within a $\pm 1\%$ tolerance, further ``burn-in'' steps are undertaken in order to thermalize the configuration.  In each step, two vortices are randomly selected and moved by different, random amounts.  If $E/N_{\varv}$ remains within the tolerance the step is accepted.  Throughout all of this process, changes that bring vortices closer than one healing length are rejected.  

To ensure that sufficient burn-in steps are used to thermalize the configurations fully we first test the effect of increasing the number of steps on the distribution of dipole lengths.  Vortices and antivortices are paired up into dipoles, starting with the two closest and continuing recursively until all are allocated to a dipole as in Ref.~\cite{Reeves2013}.  The lengths of the dipoles are then plotted as a histogram for different numbers of burn-in steps.  After a total of $\order{10^{5}}$ steps it is clear that further steps make no difference to the distribution and we consider the configuration to be thermalized.  

\begin{figure}
	\centering
	\includegraphics[width=8.7cm]{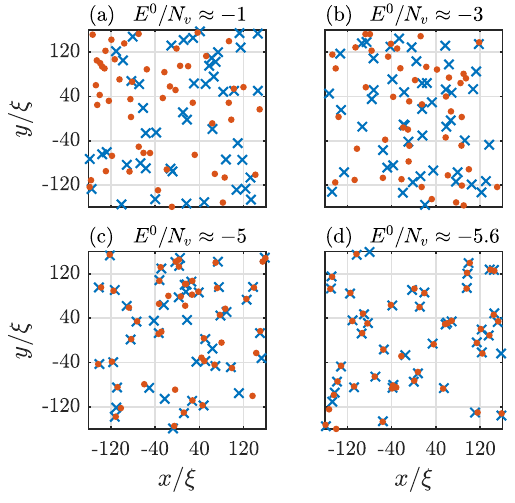}
	\caption{Examples of initial configurations of point vortices at different energies per vortex, $E^{0}/N_{\varv}$.  Blue crosses mark the positions of vortices and filled red circles mark antivortices.  From (a) to (d) the chosen value of $E^{0}/N_{\varv}$ falls further below zero and the vortices and antivortices become more closely bound into dipoles.}\label{fig:example_initial_conditions}
\end{figure}

Using this method we create ensembles of $100$ sets of configurations at initial values of energy per vortex, $E^{0}/N_{\varv}\approx\{-1,-2,-3,-4,-4.5,-5,-5.1,...,-5.6\}$.  Fig.~\ref{fig:example_initial_conditions} shows some examples of initial vortex configurations.  Note that as $E^{0}/N_{\varv}$ approaches $-5.6$ the vortices (blue crosses) and antivortices (red circles) are more closely bound into well-defined dipoles. For $E/N_{\varv}\approx -5.6$ the vortices and antivortices are so closely bound into dipoles that almost all proposed steps in the biased random walk algorithm result in increasing the energy and are therefore rejected.  This makes it impractical to create initial conditions at lower $E^{0}/N_{\varv}$.

Eq.~\ref{eqn:initial_gamma} describes a power law relationship for the initial distribution of dipole lengths for a system of independent dipoles, derived from the condition for equilibrium of the Fokker-Planck equation.  A similar relationship may also be derived using thermodynamic arguments for independent dipoles in a doubly-periodic square box and we outline this argument below.  A more detailed proof may be found in Appendix \ref{sec:appendix} at the end of this paper.

From Eq.~\ref{eqn:pv_H_periodic} it can be shown numerically that a single, isolated dipole of length $l_{i}$, contributes an energy per vortex of $E/N_{\varv}\approx C + \log{l_{i}}$, where $C$ is a constant that depends on the box dimensions (in our case $C=-6.137$).  By considering the different lengths, positions and orientations of $N_{d}$ dipoles that could yield an overall energy per vortex $E/N_{\varv}\leq\mathcal{E}$ (subject to the constraint that $l_{i}\geq 1, \forall i$) we derive an expression for the phase space volume, $\mathcal{V}(\mathcal{E})$, of such configurations.  Using the relationship $\Omega(\mathcal{E})=d\mathcal{V}/d\mathcal{E}=\beta= 1/k_{B}T$, where $k_{B}$ is the Boltzmann constant and T the temperature, and the fact that we expect the dipole lengths to follow a Boltzmann distribution, we show that:
\begin{equation}
p(l)\sim l^{\alpha}\,,\quad\text{where }\alpha=-1-\frac{1-1/N_{d}}{E/N_{\varv}+6.137}\,. \label{eqn:TB_exp_main_text}
\end{equation}
In this expression $p(l)$ is the probability distribution of dipole lengths $l$ and is related to $\Gamma(l)$ by $p(l)/l\sim\Gamma(l)$ (in order to fulfill the normalization condition in Eq.~\ref{eqn:Forrester_normalization}).  Therefore, comparing Eq.~\ref{eqn:TB_exp_main_text} with Eq.~\ref{eqn:initial_gamma}, we expect:
\begin{equation}
-2\pi K_{i}=\alpha-1\,.
\label{eqn:matching_exponents_main_text}
\end{equation}

\begin{figure}
	\centering
	\includegraphics[width=8.7cm]{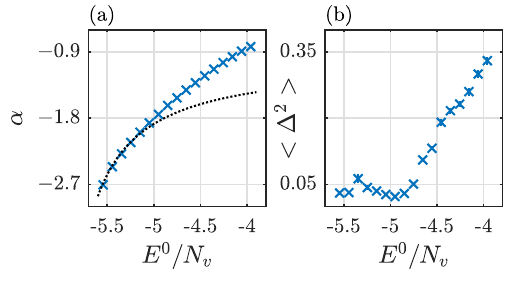}
	\caption{Comparison of initial conditions with power law predictions.  (a) The crosses indicate the mean value of the exponent, $\alpha$, from fitting a power law to $100$ bootstrap resamplings \cite{Efron1993} of the binned dipole lengths found from the initial conditions using a recursive dipole allocation algorithm.  These are shown for a selection of values of initial energy per vortex, $E^{0}/N_{\varv}$.  The uncertainties in $\alpha$, estimated from the bootstrap resampling, and the ranges in $E^{0}/N_{\varv}$, from the maximum to the minimum $E^{0}/N_{\varv}$ in the ensemble, are too small to be visible in this plot.  The dotted curve is the predicted value for $\alpha$ from Eq.~\ref{eqn:TB_exp_main_text} for fully independent dipoles.  Our initial conditions match the prediction only for $E^{0}/N_{\varv}\lesssim -5$.  (b)  The mean squared fractional difference between the power law fits and the numbers of dipoles in each bin.  The average is taken over all of the bins and the crosses and the vertical error bars indicate the means and standard deviations from $100$ bootstrap resamplings of the data.}\label{fig:fitting_to_initial_conditions}
\end{figure}

Fig.~\ref{fig:fitting_to_initial_conditions} shows how well our thermalized initial conditions of vortices and antivortices compare with these predictions for independent dipoles.  Fig.~\ref{fig:fitting_to_initial_conditions} (a) plots the exponent of the power law fit to dipole lengths (obtained by recursively allocating vortices and antivortices to dipoles from smallest to largest) against initial energy per vortex, $E^{0}/N_{\varv}$.  It is close to the prediction of Eq.~\ref{eqn:TB_exp_main_text} only for $E^{0}/N_{\varv}\lesssim -5$.  Fig.~\ref{fig:fitting_to_initial_conditions} (b) shows the squared fractional error in fitting a power law to the initial dipole distribution.  This, likewise, only shows a good fit for $E^{0}/N_{\varv}\lesssim -5$.  In short, the initial conditions we use for our point vortex modelling approximate a system of independent, well-separated dipoles only for very low energies per vortex.  Consequently, we do not expect the validity of the Fokker-Planck method to extend to cases with $E^{0}/N_{\varv}$ much greater than $-5$.   
\subsection{\label{subsec:dPVM}Dissipative Point Vortex Dynamics}
Once we have created the initial conditions, we simulate the ensuing dynamics of the vortices and antivortices using a dissipative point vortex model, with annihilations now included for vortices and antivortices that approach within one healing length of each other.  In the context of this model, our thermalized initial conditions are no longer in equilibrium and instead now evolve towards a new equilibrium state in which all vortices and antivortices have annihilated.  

Eq.~\ref{eqn:pv_eqm_periodic} gives the equations of motion for vortices in a doubly-periodic square box.  This gives us a non-dissipative velocity, $\tilde{\mathbf{v}}_{i}=(d\tilde{x}_{i}/dt,d\tilde{y}_{i}/dt)$, for each vortex.  To include the effects of dissipation we add a dissipative component at right angles to this to give the overall velocity, $\tilde{\mathbf{v}}_{i}^{\rm{diss}}=\tilde{\mathbf{v}}_{i}-\gamma\kappa_{i}\hat{\mathbf{z}}\times\tilde{\mathbf{v}}_{i}$, in which $\gamma$ is the coefficient of dissipation \cite{Billam2015}.  We use a Dormand-Prince 8th Order Runge-Kutta algorithm \cite{Prince1981} to solve these equations of motion, with an absolute tolerance of $1\times 10^{-8}$.  We simulate the annihilation of vortices and antivortices that come into too close proximity by removing those that come closer than one healing length, which in these units is $\tilde{\xi}=2\pi/L$.  Finally, we rescale lengths back to units of $\xi$ and rescale times such that $\kappa=\pm2\pi$.  Note that in order to achieve the wholy dissipative case we set $\gamma=1$ to calculate $\tilde{\mathbf{v}}_{i}^{\rm{diss}}$ and then remove the non-dissipative component of velocity.

For a given initial energy per vortex, $E^{0}/N_{\varv}$, and dissipation, $\gamma$, we carry out the above procedure for each of the $100$ initial conditions in our ensemble, saving the vortex coordinates and signs at regular time-steps.  We then allocate all of the vortices and antivortices into dipoles, recursively from smallest to largest, as in Sec.~\ref{subsec:initial_conditions}.  We do this at each time step and save the dipole lengths.  The result, for each value of $E/N_{\varv}$ and $\gamma$, is an ensemble of $100$ sets of point vortex coordinates and corresponding values for the dipole lengths, as they evolve over time in accordance with the dissipative point vortex model.

\begin{figure*}
	\centering
	\includegraphics[width=17.4cm]{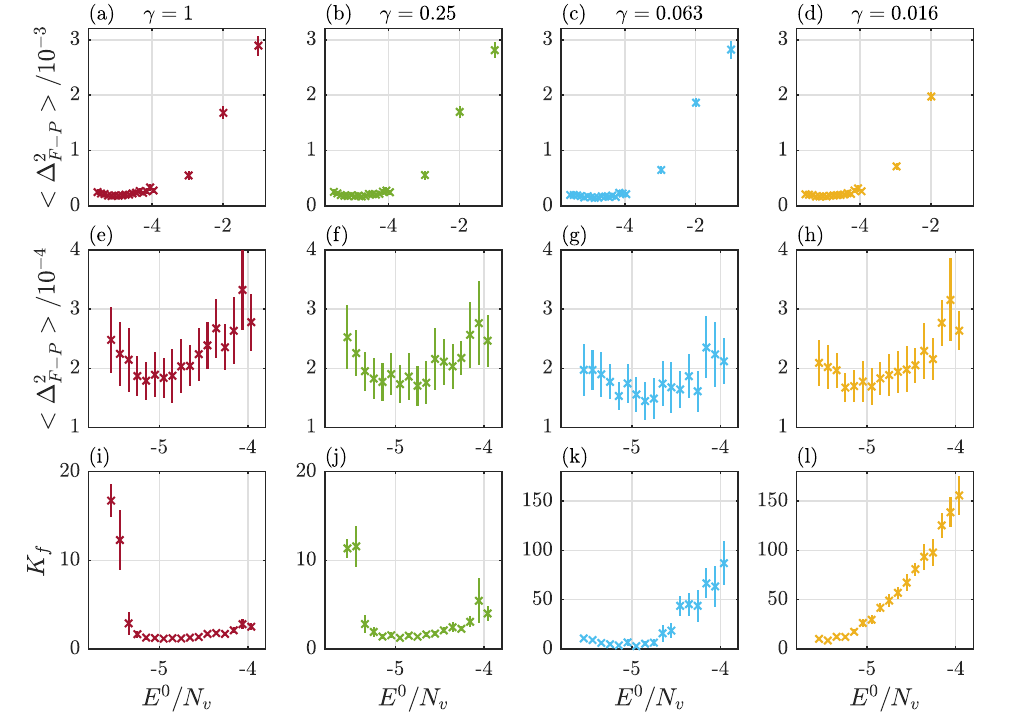}
	\caption{Comparison of the probability distribution for dipole lengths using the Fokker-Planck equation, $\Gamma_{F}$, with the results of dissipative point vortex modelling, $\Gamma_{P}$. Top row, (a)-(d): an average over all lengths and times of the squared fractional differences between the logarithms of $\Gamma_{F}$ and $\Gamma_{P}$, $<\Delta_{F-P}^{2}>$, plotted against initial energy per vortex, $E^{0}/N_{\varv}$, for four different levels of the dissipation, $\gamma=\{1,2^{-2},2^{-4},2^{-6}\}$.  Middle row, (e)-(h): as for the top row but just showing $E^{0}/N_{\varv}\lesssim-4$.  Bottom row, (i)-(l): the values of $K_{f}$ required in the Fokker-Planck equation to yield the minimum values of $<\Delta_{F-P}^{2}>$ plotted in the middle row.  In all plots, the crosses indicate the means and the vertical error bars the standard deviations in values calculated across $100$ bootstrap resamplings.  Horizontal errorbars (not visible) show the full range of $E^{0}/N_{\varv}$ in each ensemble. }\label{fig:comparison_graphs}
\end{figure*}

\subsection{\label{subsec:FPE}Fokker-Planck Evolution}
To model Fokker-Planck evolution for a given $E^{0}/N_{\varv}$ and $\gamma$, we first take the initial dipole lengths from the corresponding point vortex results and construct an initial probability distribution, $\Gamma(l,0)$, by binning the lengths from all $100$ realizations in an ensemble into $480$ equally-sized bins \footnote{To determine a suitable number of bins we increased the number used until numerical converegence was observed.}, starting at $l=1$.  We then normalize using the condition in Eq.~\ref{eqn:Forrester_normalization}, where $N_{d}$ is the initial number of dipoles.  Using this initial probability distribution of dipole lengths, we numerically model the Fokker-Planck evolution described by Eq.~\ref{eqn:FP_dipoles_Forrester}.  We employ a $2^{\text{nd}}$ order central-difference scheme for the spatial derivatives in the bulk, a forward-difference scheme at the lower, $l=1$, boundary and at the upper boundary we use a central-difference scheme with a ``ghost'' point beyond the boundary calculated to ensure continuity of flux \cite{Chu2001}.  We combine this with a $4^{\text{th}}$ order Runge-Kutta method, with a relative tolerance of $1\times 10^{-3}$, for the temporal evolution \cite{Dormand1980,Prince1981}.  We use absorbing boundary conditions \cite{Gardiner2009} to model the annihilation of dipoles whose length drifts below $l=1$.  Consequently, the number density of dipoles, $\rho$, found using the integral in Eq.~\ref{eqn:Forrester_normalization}, decreases over time. 

The final dimensionless superfluid density that we are quenching to in the Fokker-Planck method, $K_{f}$, is not a parameter in the point vortex model.  Therefore, we treat is as an undetermined parameter to be determined by fitting.  We test a range of values of $K_{f}$ and select the one which gives the closest match to the point vortex results.  This comparison is measured between $\Gamma_{F}(l,t)$, the distribution of dipole lengths over time from the Fokker-Planck evolution and $\Gamma_{P}(l,t)$, the distribution of dipole lengths found by binning the lengths for the whole ensemble of $100$ point vortex data sets, for a given $E^{0}/N_{\varv}$ and $\gamma$.  From these we calculate the the squared fractional differences between the logarithms of $\Gamma_{F}(l,t)$ and $\Gamma_{P}(l,t)$, and average over all available dipole lengths and simulation times:
\begin{equation}
<\Delta_{F-P}^{2}>=\Biggl<\left(\frac{\log_{10}\Gamma_{P}-\log_{10}\Gamma_{F}}{\log_{10}\Gamma_{F}}\right)^{2}\Biggr>\,.
\label{eqn:comparison_metric}
\end{equation}
We then select the value of $K_{f}$ that minimizes the comparison metric $<\Delta_{F-P}^{2}>$.

To estimate the variation in these values of $K_{f}$ and $<\Delta_{F-P}^{2}>$, we employ a bootstrapping technique \cite{Efron1993} when sampling the initial point vortex dipole lengths to create the Fokker-Planck initial distribution $\Gamma_{F}(l,0)$.  Rather than use each of the $100$ sets of initial dipole lengths once, we instead sample these $100$ sets randomly $100$ times with replacement.  We use this sample to create $\Gamma_{F}(l,0)$.  After completing the analysis described above we then repeat, randomly resampling the initial point vortex dipole lengths to create another $\Gamma_{F}(l,0)$ and so on.  We then repeat this process $100$ times,  with a different resampling of the initial conditons each time, and take the mean and standard deviations of the values of $K_{f}$ and $<\Delta_{F-P}^{2}>$ we find. 

\section{\label{sec:results}Results and Analysis}
\begin{figure*}
	\centering
	\includegraphics[width=17.4cm]{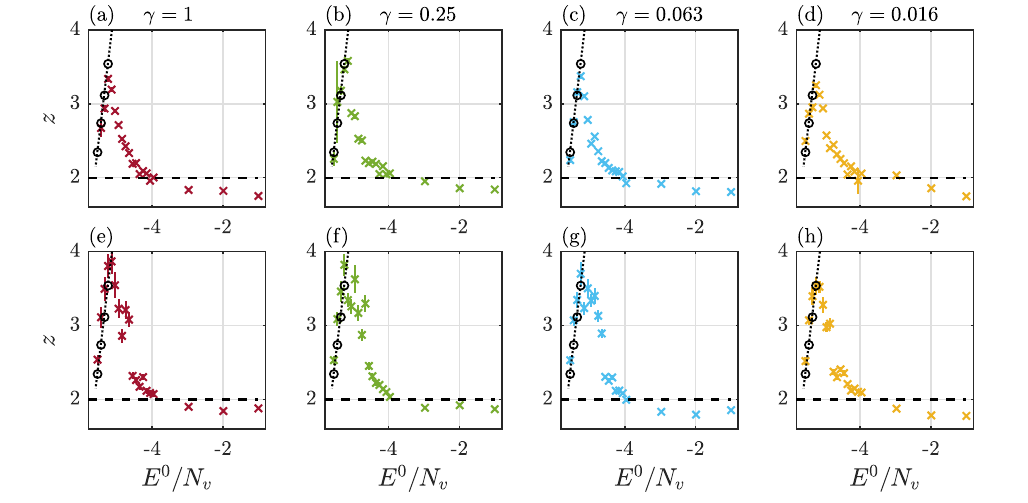}
	\caption{The dependence of the dynamical critical exponent, $z$, on initial energy per vortex, $E^{0}/N_{\varv}$, for four different levels of dissipation, $\gamma=\{1,2^{-2},2^{-4},2^{-6}\}$.  Top row, (a)-(d): The crosses and vertical error bars represent means and standard deviations in $z$ found by bootstrap resampling $L_{c}(t)$ $100$ times, conducting a power law fit each time and finding the mean and standard deviation in $z$ across the resamplings.  This is done only using time-steps for which scaling has been demonstrated by the collapse of the correlation function.  Bottom row, (e)-(h): As for the top row except that $z$ in these cases is determined using the less reliable method of fitting a power law to the number of vortices of the form $N_{\varv}\sim t^{-2/z}$ and finding the mean and vertical error bars for $z$ from $100$ bootstrap resamplings.  The window used for the fitting was adjusted to achieve the closest fit with the data and the values of $z$ plotted are for this window.  In all plots the horizontal errorbars (not visible) show the full range of $E^{0}/N_{\varv}$ in each ensemble.  The dashed horizontal line indicates $z=2$ and the circles indicate the predicted value of $z$ from Eq.~\ref{eqn:FP_z} where $2\pi K_{i}$ is found from the magnitude of the exponent in a power fit to the initial dipole lengths (see Fig.~\ref{fig:fitting_to_initial_conditions}).  The dotted line shows the predicted value for $z$ using Eq.~\ref{eqn:FP_z} where $2\pi K_{i}$ is found using the exponent $\alpha$ from Eq.~\ref{eqn:TB_exp_main_text} and then using Eq.~\ref{eqn:matching_exponents_main_text}. }\label{fig:scaling_graphs}
\end{figure*}
Fig.~\ref{fig:comparison_graphs} shows a selection of results comparing the evolution of the dipole length distributions using the dissipative point vortex model with the evolution using the Fokker-Planck equation.  Four levels of dissipation are plotted, with the top row showing how the comparison metric defined in Eq.~\ref{eqn:comparison_metric} depends on the initial energy per vortex, $E^{0}/N_{\varv}$, across a range from $-1$ down to $-5.6$.  It is clear that the Fokker-Planck approach closely matches the point vortex results only for a limited range of $E^{0}/N_{\varv}$, below about $-4$, and that this is a consistent pattern across different levels of dissipation.  The middle row shows the same data for just the region below $E^{0}/N_{\varv}\approx-4$ and suggests that $E^{0}/N_{\varv}\approx-5$ may give the closest comparison.  It seems likely that part of the reason for the failure at higher $E^{0}/N_{\varv}$ is because the point vortex configurations cannot be safely treated as a system of independent dipoles (see Fig.~\ref{fig:fitting_to_initial_conditions}).

The bottom row of Fig.~\ref{fig:comparison_graphs} shows the value of $K_{f}$ needed in the Fokker-Planck method to give the closest possible fit to the point vortex results.  Here there is a difference in behaviour at different levels of dissipation, $\gamma$.  For higher $\gamma$, $K_{f}$ has a  consistent, low value across a range of $E^{0}/N_{\varv}$ from $-4$ down to about $-5.3$.  However, at lower dissipation $K_{f}$ varies widely with a large increase in value as $E^{0}/N_{\varv}$ increases above $-5$.  

The role of $K_{f}$ in the Fokker-Planck equation (see Eq.~\ref{eqn:FP_dipoles_Forrester}) is to regulate the ratio of the drift term, associated with the shrinking of dipoles due to dissipation, to the diffusion term, associated with random variations in dipole length.  High values of $K_{f}$ imply that drift dominates diffusion.  In our dissipative point vortex model there are no thermal effects, therefore random variations in dipole lengths can only be due to interactions with the vortices in other dipoles.  The very high values of $K_{f}$ for some lower dissipation results suggest that a pure drift in dipole lengths is as good a fit to the dissipative point vortex model as can be achieved.  Note that higher values of $K_{f}$ are required for all dissipations at the very lowest $E^{0}/N_{\varv}$, suggesting that inter-dipole interactions become less significant as vortices are bound ever more tightly into their dipole pairs.  In essence, a good fit between the Fokker-Planck and dissipative point vortex methods is seen at $E^{0}/N_{\varv}\lesssim -4$, but this comparison is less convincing at lower levels of dissipation.  The initial configurations must be sufficiently dipole dominated for the Fokker-Planck method to apply and the dissipation must be high enough to shrink and annihilate the dipoles whilst interactions with other dipoles or free vortices remain relatively small and random in nature.

To test the comparison of the dissipative point vortex results with the Fokker-Planck model further, we investigate the evidence for dynamical scaling in the point vortex data and calculate the dynamical critical exponent $z$.  Following a process detailed in \cite{Tattersall2024}, the fluid velocity, $\mathbf{v}$, is reconstructed on a grid using the point vortex coordinates and signs at each time-step.  From this, a two-point correlation function is found based on the corrected speed, $v_{c}=|\mathbf{v}|-\langle |\mathbf{v}| \rangle$, where the angled brackets represent an average over all grid-points.  We then search for evidence of scaling in the time dependence of this function.  Scaling is agreed to hold if the correlation functions at different time-steps collapse to a single curve once rescaled by $L_{c}(t)$, as described in Eq.~\ref{eqn:scaling}.  In our case, we take $L_{c}(t)$ to be the distance at which the correlation function falls to $0.1$ from a value of one at $r=0$.  A systematic test for this collapse, by measuring the similarity of the rescaled correlation functions using the Mahalanobis distance \cite{Mahalanobis1936_2018} is used.  Once the scaling region is identified in this manner, we conduct a least squares fit of the power law $L_{c}\sim t^{1/z}$ to the data, for just the times within the scaling region.  This approach has the benefit, over simply fitting a power law to the time dependence of vortex number (or density), in that we first verify that dynamical scaling holds for a given time window before conducting our fit to determine $z$. 
\begin{figure*}
	\centering
	\includegraphics[width=17.4cm]{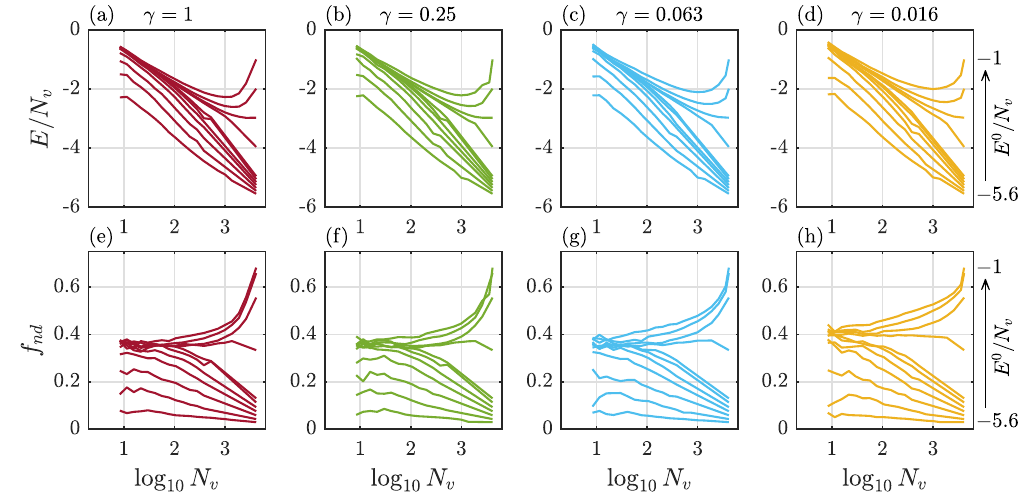}
	\caption{Tracking the variation in vortex configurations as the number of vortices remaining decreases.  Top row, (a)-(d): The variation in energy per vortex, $E/N_{\varv}$, with number of vortices, $N_{\varv}$.  Each line plotted shows the variation in $E/N_{\varv}$ from a given initial value, $E^{0}/N_{\varv}\approx\{-5.6,-5.5,-5.4,-5.3,-5.2,-5.1,-5,-4,-3,-2,-1\}$ on the right, to late times, when the number of vortices has fallen significantly, on the left.  The lines show the mean value of $E/N_{\varv}$ calculated across all time-steps in all $100$ runs in the ensemble which have $N_{\varv}$ in a given range.  The standard errors in $E/N_{\varv}$ and $N_{\varv}$ were calculated but were smaller than the thickness of the line.  Each subplot (a)-(d) corresponds to a different level of dissipation $\gamma=\{1,2^{-2},2^{-4},2^{-6}\}$.  Bottom row, (e)-(h): The variation in the fraction of vortices not allocated to dipoles, $f_{nd}$, calculated using the algorithm described in the final paragraph of section~\ref{sec:results}.  The same initial energies per vortex, $E^{0}/N_{\varv}$, are plotted as in (a)-(d) and start on the right of each plot in the same order (lower $E^{0}/N_{\varv}$ also has lower initial $f_{nd}$).  In all eight plots, results for $E^{0}/N_{\varv} > -5.2 $ converge as $N_{\varv}$ decreases whilst those for $E^{0}/N_{\varv}\lesssim-5.2 $ remain distinct.  }\label{fig:WM_graphs_2}
\end{figure*}

Fig.~\ref{fig:scaling_graphs} (a)-(d) shows the values of $z$ calculated in this manner for four different dissipations, plotted against $E^{0}/N_{\varv}$.  Fig.~\ref{fig:scaling_graphs} (e)-(f) shows the same plots but with $z$ calulated using a less reliable method that finds the window at which $N_{\varv}\sim t^{-2/z}$ best fits the data and extracting $z$ from the mean of $100$ bootstrap resamplings of this fit.  It is clear from both versions of these plots that above $E^{0}/N_{\varv}\approx-4$ the data broadly agree with the Model A prediction of $z=2$ \cite{Hohenberg1977}.  However, as $E^{0}/N_{\varv}$ falls further below this value, $z$ increases rapidly until it meets, and then follows, the dotted line corresponding to Eq.~\ref{eqn:FP_z}.  This is the prediction from \cite{Forrester2013} for the Fokker-Planck model.  This crossover in behaviour of the dynamical critical exponent is consistent across all values of dissipation tested.  Note that in order to extract a value for $K_{i}$ in Eq.~\ref{eqn:FP_z} we use the value of $\alpha$ from a power law fit to the initial conditions (see Fig.~\ref{fig:fitting_to_initial_conditions}) and Eq.~\ref{eqn:matching_exponents_main_text}.

To investigate this crossover in more detail we calculate the energy per vortex, $E/N_{\varv}$, at \textit{all} time-steps in the point vortex data.  The results of this are plotted in Fig.~\ref{fig:WM_graphs_2} (a)-(d) against the number of vortices remaining, for four different levels of dissipation.  These graphs show that point vortex simulations across a wide range of initial conditions, down to $E^{0}/N_{\varv}\approx-5.2$, converge towards similar energies per vortex as the vortex number decays.  This ``forgetting'' of initial conditions is typical of universal scaling, with trajectories from multiple starting points converging on similar end points at late times.  However, for $E^{0}/N_{\varv}\lesssim-5.2$, this convergence no longer occurs, with each trajectory remaining distinct throughout.  This represents additional evidence of a change in scaling behaviour of the dissipative point vortex model for very low initial energies per vortex. 

To test if this lack of convergence in $E/N_{\varv}$ is related to the prevalence of well-defined dipoles in the vortex configurations, we employ an algorithm that allocates vortices and antivortices into either clusters, dipoles or neither.  At each time-step for each run of the point vortex results, this algorithm first searches for clusters of only vortices or only antivortices using a density based scanning approach (DBSCAN) \cite{Ester1996}.  Like-signed vortices that are closer than the average intervortex distance, and have no other opposite-signed vortices nearby, are allocated to clusters so long as there are at least two of them.  The remaining vortices are then searched for evidence of dipoles.  A vortex and an antivortex are allocated to a dipole if they are closer than $0.8\,\times\,$the average intervortex distance, are each others nearest-neighbour and their next-nearest neighbours are no closer than $1.4\,\times\,$the dipole length.  All vortices not allocated to clusters or dipoles count as ``neither''.  The fraction of all vortices not allocated to dipoles, $f_{nd}$, is plotted against number of vortices remaining, $N_{\varv}$, in Fig.~\ref{fig:WM_graphs_2} (e)-(f), for four different levels of dissipation.  The lower the initial energy per vortex, $E^{0}/N_{\varv}$, the lower the initial value of $f_{nd}$ on the right of each graph.  Just as was seen in the $E/N_{\varv}$ plots, initial conditions down to $E^{0}/N_{\varv}\approx-5.2$ converge (to $f_{nd}\approx 0.4$) at late times.  But those with $E^{0}/N_{\varv}\lesssim-5.2$ maintain higher dipole fractions throughout and do not converge.

\section{\label{sec:conclusions}Conclusions}
In this paper we have presented numerical evidence that modelling coarsening behaviour in two-dimensional superfluids with a dissipative point vortex model tends towards the results of a Fokker-Planck equation at $E^{0}/N_{\varv}\lesssim -4$.  We suggest that this is due to the vortices and antivortices being sufficently closely bound into dipoles that the ensuing dynamics mimics the drift towards shorter dipole lengths, with some random variation, that is encoded in the Fokker-Planck equation.  The comparison holds better for results at higher levels of dissipation, in which dipoles tend to shrink and annihilate before any more complex dynamics, caused by close interaction with other dipoles or vortices, can occur.  For $E^{0}/N_{\varv}>-4$ the dissipative point vortex model and the Fokker-Planck equation no longer agree.

We also observed a crossover in the behaviour of the dynamical critical exponent, $z$, determined from the point vortex data.  For $E^{0}/N_{\varv}>-4$ we find a value of $z$ that is close to two and independent of initial conditions, in line with expectations for Model A universal dynamical scaling.  Whereas for  $E^{0}/N_{\varv}\lesssim-5$, $z$ depends explicitly on the initial conditions, as predicted using the Fokker-Planck method.  This crossover is consistent across all levels of dissipation tested.  Furthermore, we have observed that for $E^{0}/N_{\varv}\gtrsim-5.2$ all initial conditions converge to similar values of $E/N_{\varv}$ and fraction of non-dipole vortices at late times but those with $E^{0}/N_{\varv}\lesssim-5.2$ do not.  This suggests that those configurations that are most tightly bound into dipoles at the beginning never fully ``forget'' their initial conditions.

Note that the exact position of this crossover should not be considered to be universal.  We had a fixed initial density of vortices for our simulations and it may be that the position of the crossover is sensitive to changes in this density.  We have not investigated initial energies per vortex greater than zero, which tend towards clustering of like-signed vortices, as it is clear that the comparison with the Fokker-Planck description breaks down well before this.  However, it would be interesting to know if the convergence of $E/N_{\varv}$, witnessed in Fig.~\ref{fig:WM_graphs_2}, for $-5.2\lesssim E^{0}/N_{\varv}\lesssim -1$ is maintained for highly clustered initial configurations with $E^{0}/N_{\varv}>>0$.

\section{Acknowledgements}

\noindent This work was supported by a Lady Bertha Jeffreys PhD Studentship.\\

\noindent RT thanks G. A. Williams for valuable correspondence and input on dynamical scaling in quenched two-dimensional superfluids.\\

\noindent For the purpose of open access, the author has applied a Creative Commons Attribution (CC BY) licence to any Author Accepted Manuscript version arising from this submission.

\bibliography{FP_PVM_Refs}

\clearpage
\appendix

\section{\label{sec:appendix}Proof of Power Law for Initial Dipole Lengths by Thermodynamic Arguments}
Here we present, in more detail, a proof of Eq.~\ref{eqn:TB_exp_main_text} from section~\ref{subsec:initial_conditions} above.  The system under consideration is one of independent, well-separated, vortex-antivortex dipoles. 

Using Eq.~\ref{eqn:pv_H_periodic} it can be demonstrated numerically that the energy per vortex of a single isolated dipole of length $l$ in a doubly-periodic square box of length $L$ is well-approximated by:
\begin{align*}
E/N_{\varv}&\approx\log{\frac{2\pi a l}{L}}\,,\\
E/N_{\varv}&\approx\log{\frac{2\pi a}{L}}+\log{l}\,,
\label{eqn:TBE_single}
\end{align*}
where $a$ is determined to be $0.705$ for a box of length $L=2048$.  This can be re-written as:
\begin{equation}
E/N_{\varv}\approx -6.137+\log{l}\,.
\label{eqn:TBE_single_a}
\end{equation}
It follows that for a system of $N_{d}$ well separated dipoles of lengths $l_{i}$, the energy per vortex is given by:
\begin{equation}
E/N_{\varv}\approx -6.137+\frac{1}{N_{d}}\sum_{i=1}^{N_{d}} \log{l_{i}}\,.
\label{eqn:TBE_2}
\end{equation} 
Note that Eq.~\ref{eqn:TBE_2} implies that a surface of constant $E/N_{\varv}$, in an $N_{d}$-dimensional phase space, in which the dipole lengths define the coordinates of a point, obeys the condition:
\begin{equation}
\sum_{i=1}^{N_{d}}\log{l_{i}}=\text{constant}\,.
\label{eqn:constant_E_condition}
\end{equation}
For a given $E/N_{\varv}=\mathcal{E}$, any single configuration of dipoles can be entirely described by their lengths, $l_{i}$, their orientations, $\theta_{i}$ and their centre of mass coordinates, $(x_{i},y_{i})$:
\begin{equation*}
0\leq x_{i}<L\,,\quad 0\leq y_{i}<L\,,\quad 0\leq \theta_{i}<2\pi\,,\quad l_{min}\leq l_{i}\leq l_{max}\,.
\end{equation*}
Here $l_{min}$ is the shortest dipole length and $l_{max}$ is the longest possible dipole length, which occurs when every other dipole has length $l_{min}$.  Consequently $l_{max}$ is related to $\mathcal{E}$, $N_{d}$ and $l_{min}$ via:
\begin{align}
\mathcal{E}&\approx -6.137+\frac{N_{d}-1}{N_{d}}\log{l_{min}}+\frac{1}{N_{d}}\log{l_{max}}\,,\nonumber\\
\mathcal{E}&\approx -6.137+\log{l_{min}}+\frac{1}{N_{d}}\log{\frac{l_{max}}{l_{min}}}\,.
\label{eqn:l_max_1}
\end{align}
Taking Khinchin's approach to microcanonical statistics \cite{Khinchin1949}, the phase space volume, $\mathcal{V}$, with $E/N_{\varv}\leq\mathcal{E}$ is given by:
\begin{align}
\mathcal{V}(\mathcal{E})&=\int_{E/N_{\varv}\leq\mathcal{E}}dV \nonumber\\
&=\int_{E/N_{\varv}\leq\mathcal{E}}\prod_{i=1}^{N_{d}}dx_{i}dy_{i}l_{i}dl_{i}d\theta_{i} \nonumber\\
&= (2\pi)^{N_{d}}L^{2N_{d}}\int_{E/N_{\varv}\leq\mathcal{E}}\prod_{i=1}^{N_{d}}l_{i}dl_{i}\,.
\label{eqn:phase_space_volume}
\end{align}
Consider the quantity:
\begin{equation*} \mathcal{L}_{i}=\log{l_{i}/l_{min}}=\log{l_{i}}-\log{l_{min}}\,. 
\end{equation*} 
It follows that:
\begin{equation*}
dl_{i}=l_{i}d\mathcal{L}_{i}\,\,\text{and}\,\, l_{i}=l_{min}\exp{\mathcal{L}_{i}}\,.
\end{equation*}
Substituting these into Eq.~\ref{eqn:phase_space_volume} gives:
\begin{equation}
\mathcal{V}(\mathcal{E})=(2\pi)^{N_{d}}L^{2N_{d}}l_{min}^{2N_{d}}\int_{E/N_{\varv}\leq\mathcal{E}}\exp{2\sum_{i=1}^{N_{d}}\mathcal{L}_{i}}\prod_{i=1}^{N_{d}}d\mathcal{L}_{i}\,.
\label{eqn:phase_space_vol_1}
\end{equation}
Furthermore, the condition for a surface of constant $E/N_{\varv}$ in Eq.~\ref{eqn:constant_E_condition} becomes:
\begin{equation}
\sum_{i=1}^{N_{d}}\mathcal{L}_{i}=\text{constant}\,.
\end{equation}
and, using $\mathcal{L}=\log{l_{max}/l_{min}}$, Eq.~\ref{eqn:l_max_1} becomes:
\begin{equation}
\mathcal{L}_{max}\approx(\mathcal{E}+6.137-\log{l_{min}})N_{d}\,.
\label{eqn:L_max_and_E}
\end{equation}
For $N_{d}$ dipoles the ``volume'' for the integral in Eq.~\ref{eqn:phase_space_vol_1} is that of a simplex with an orthogonal corner and side length $\mathcal{L}_{max}$.  Hence $V=\mathcal{L}_{max}^{N_{d}}/(N_{d}!)$ and the ``surfaces'' correspond to $\sum_{i=1}^{N_{d}}\mathcal{L}_{i}=\text{constant.}$  It follows that:
\begin{equation*}
\frac{dV}{d\mathcal{L}_{max}}=\frac{\mathcal{L}_{max}^{N_{d}-1}}{(N_{d}-1)!}\,.
\end{equation*}
If we introduce the dummy variable $\mathcal{L}'= \sum_{i=1}^{N_{d}}\mathcal{L}_{i}$, where $0\leq\mathcal{L}'\leq\mathcal{L}_{max}$ and each value of $\mathcal{L}'$ corresponds to a surface of constant $E/N_{\varv}$, then the volume element for the integral in Eq.~\ref{eqn:phase_space_vol_1} can be replaced with:
\begin{equation*}
dV=\frac{\mathcal{L}'^{N_{d}-1}d\mathcal{L}'}{(N_{d}-1)!}\,,
\end{equation*}
and the integral can be re-written as:
\begin{equation}
\mathcal{V}(\mathcal{E})=(2\pi)^{N_{d}}L^{2N_{d}}l_{min}^{2N_{d}}\int_{\mathcal{L}'=0}^{\mathcal{L}'=\mathcal{L}_{max}}\exp{2\mathcal{L}'}\frac{\mathcal{L}'^{N_{d}-1}d\mathcal{L}'}{(N_{d}-1)!}\,.
\label{eqn:TB_integral_changed_variables}
\end{equation}
From the fundamental theorem of calculus, it follows that:
\begin{equation*}
\frac{d\mathcal{V}}{d\mathcal{L}_{max}}=(2\pi L^{2} l_{min}^{2})^{N_{d}}\exp{2\mathcal{L}_{max}}\frac{\mathcal{L}_{max}^{N_{d}-1}}{(N_{d}-1)!}\,.
\end{equation*}
We are interested in the quantity $\Omega(\mathcal{E})=d\mathcal{V}/d\mathcal{E}$, since $d\log{\Omega}/d\mathcal{E}=\beta=1/k_{B}T$.  Using the chain rule:
\begin{equation*}
\Omega(\mathcal{E})=\frac{d\mathcal{V}}{d\mathcal{L}_{max}}\frac{d\mathcal{L}_{max}}{d\mathcal{E}}\,,
\end{equation*} 
and noting that, from Eq.~\ref{eqn:L_max_and_E}, $d\mathcal{L}_{max}/d\mathcal{E}=N_{d}$, we get:
\begin{equation*}
\Omega(\mathcal{E})=N_{d}(2\pi L^{2} l_{min}^{2})^{N_{d}}\exp{2\mathcal{L}_{max}}\frac{\mathcal{L}_{max}^{N_{d}-1}}{(N_{d}-1)!}\,.
\label{eqn:omega}
\end{equation*}
Hence:
\begin{align*}
\log{\Omega}=\log{N_{d}}+&N_{d}\log(2\pi L^{2} l_{min}^{2})+2\mathcal{L}_{max}\\ &+(N_{d}-1)\log{\mathcal{L}_{max}}-\log{(N_{d}-1)!}\,,
\end{align*} 
and so:
\begin{equation}
\beta=\frac{d\log{\Omega}}{d\mathcal{E}}=\frac{d\log{\Omega}}{d\mathcal{L}_{max}}\frac{d\mathcal{L}_{max}}{d\mathcal{E}}=\left(2+\frac{N_{d}-1}{\mathcal{L}_{max}}\right)N_{d}\,.
\end{equation}
Using Eq.~\ref{eqn:L_max_and_E} to replace $\mathcal{L}_{max}$ with an expression containing the energy per vortex $\mathcal{E}$ gives:
\begin{equation*}
\beta=2N_{d}+\frac{N_{d}-1}{\mathcal{E}+6.137-\log{l_{min}}}\,.
\label{eqn:beta}
\end{equation*}
This equation gives an unambiguous definition of a thermodynamic temperature for a system of point vortex dipoles in a doubly-periodic square box of length $L=2048$.  It simplifies further in our case as we take $l_{min}=1$ (since a vortex and antivortex annihilate at a distance of one healing length), hence:
\begin{equation}
\beta=2N_{d}+\frac{N_{d}-1}{\mathcal{E}+6.137}\,.
\label{eqn:beta_simpler}
\end{equation}
This definition fails if $\mathcal{E}$ falls to $-6.137$; this is because all of the dipoles would be of length one at this point and would annihilate. 

To get a distribution of dipole lengths we note that from Eq.~\ref{eqn:TBE_single_a} that each dipole contributes an energy per vortex of:
\begin{equation*}
\varepsilon_{i}=\frac{1}{N_{d}}(-6.137+\log{l_{i}})\,,
\end{equation*}
and has a degeneracy given by the product of the area its centre could occupy with the circumference of the circle described by all $2\pi$ possible orientations, i.e.:
\begin{equation*}
g_{i}=L^{2}\times 2\pi l_{i}=2\pi L^{2}l_{i}\,.
\end{equation*}
We expect the probability distribution of dipole lengths to go as $p(l_{i})\sim g_{i}\exp{-\beta \varepsilon_{i}}$.  From this, following a few lines of algebra, it can be shown that:
\begin{equation}
p(l_{i})\sim l_{i}^{\left(1-\frac{\beta}{N_{d}}\right)}\,.
\end{equation}
In other words, $p(l)\sim l^\alpha$, with the exponent, $\alpha$, given by $1-\beta/N_{d}$.  Substituting the expression for $\beta$ from Eq.~\ref{eqn:beta_simpler} gives:
\begin{equation}
\alpha=1-2-\frac{N_{d}-1}{N_{d}(\mathcal{E}+6.137)}=-1-\frac{1-1/N_{d}}{\mathcal{E}+6.137}\,. \label{eqn:TB_exp_proved}
\end{equation}

Note that the probability distribution $p(l)$ is related to Fokker-Planck distribution $\Gamma(l)$, from Eq.~\ref{eqn:FP_dipoles_Forrester}, by $p(l)/l\sim\Gamma(l)$, in order to fulfill the normalization condition in Eq.~\ref{eqn:Forrester_normalization}.  Therefore we expect:
\begin{equation}
-2\pi K_{i}=\alpha-1\,.
\label{eqn:matching_exponents}
\end{equation}

\end{document}